\documentclass[
  aps,
  pra,
  reprint,
  amsmath,
  amssymb,
  showpacs,
]{revtex4-1}

\usepackage{graphicx}
\usepackage{hyperref}

\newcommand{\Chi}{\mathrm{X}}

\newcommand{\tmmathbf}[1]{\ensuremath{\boldsymbol{#1}}}
\newcommand{\tmop}[1]{\ensuremath{\operatorname{#1}}}

\begin{document}

\title{Stability of Emergent Kinetics in Optical Lattices with Artificial Spin-Orbit Coupling}

\author{Mengsu Chen}

\affiliation{Department of Physics, Virginia Tech, Blacksburg, Virginia
24061, USA}

\author{V. W. Scarola}

\affiliation{Department of Physics, Virginia Tech, Blacksburg, Virginia
24061, USA}

\begin{abstract}
Artificial spin-orbit coupling in optical lattices can be engineered to tune band structure into extreme regimes where the single-particle band flattens leaving only inter-particle interactions to define many-body states of matter.   Lin et al. [Phys. Rev. Lett \textbf{112}, 110404 (2014)] showed that under such conditions interactions lead to a Wigner crystal of fermionic atoms under approximate conditions: no bandwidth or band mixing.  The excitations were shown to possess emergent kinetics with fractionalized charge derived entirely from interactions.  In this work we use numerical exact diagonalization to study a more realistic model with non-zero bandwidth and band mixing.  We map out the stability phase diagram of the Wigner crystal.  We find that emergent properties of the Wigner crystal excitations remain stable for realistic experimental parameters.  Our results validate the approximations made by Lin et al. and define parameter regimes where strong interaction effects generate emergent kinetics in optical lattices.   
\end{abstract}

{\maketitle}

\section{Introduction}

Precise control over the band structure of ultracold atoms and molecules placed in optical lattices enables access to strongly correlated states \cite{jaksch:1998,greiner:2002,bloch:2008}.  Recent work shows that optical lattices allow further exploration of extreme regimes of strong correlation where the single-particle dispersion can be flattened to emphasize interactions, much like the lowest Landau level in the fractional quantum Hall regime \cite{laughlin:1983a}.  Examples in the optical lattice context include flat single-particle bands in triangular \cite{huber:2010,becker:2010}, honeycomb \cite{wu:2007,soltan-panahi:2011,tarruell:2012}, and kagome \cite{huber:2010,jo:2012,parameswaran:2013} optical lattices.  Another example includes flat bands generated from artificial spin-orbit coupling (SOC)  \cite{higbie:2002,lin:2011,sau:2011,ramachandhran:2012,hu:2012,zhou:2013,galitski:2013} in one dimensional chains \cite{zhang:2013} and the two dimensional square lattice \cite{hui:2016}.  In all of these cases, there is an opportunity for emphasized interaction effects to lead to emergent physics wherein interactions operating within the flat band generate entanglement, as in the fractional quantum Hall regime \cite{laughlin:1983b,jain:1989,jain:2007}.

In general, interacting flat band models are captured by two distinct classes of Hamiltonian that lead to either quantum or classical states \cite{hui:2016}. 
Classical flat band models are defined by only diagonal interaction terms in a site basis.  They are trivial and exhibit only classical (unentangled) configurations of particles in the absence of kinetics because the single-particle basis states are highly localized and cannot overlap via interactions.  Examples of classical flat band problems include basic Hubbard models \cite{jaksch:1998} of atoms in very deep optical lattices without applied fields.   But in quantum flat band models, interactions are off-diagonal in a site basis and entangle particles even in the absence of any dispersive single-particle bands because the single-particle basis states are only quasi-localized and can effectively overlap via interactions.  Recent work showed that flat SOC bands in optical lattices define quantum flat bands \cite{lin:2014,hui:2016}.  
\begin{figure}
  \includegraphics[width=0.85\columnwidth]{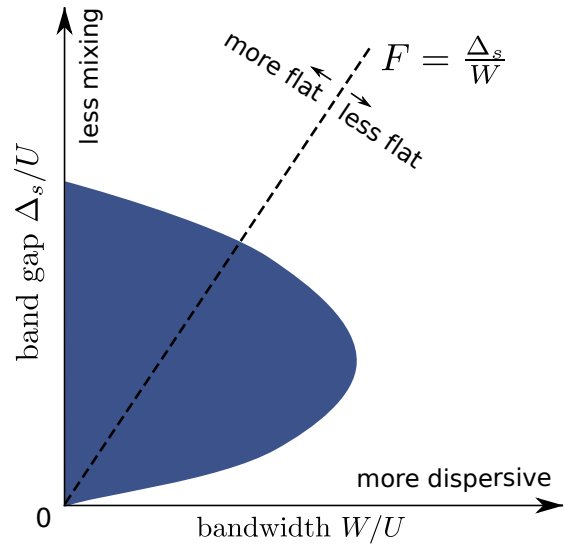}
 \caption{\label{fig_schematic}
 Schematic plotting the stability of the Wigner crystal phase (lobe) in the parameter space of single-particle band gap versus bandwidth.  $F$ defines the flatness ratio.  Increasing the single-particle bandwidth makes the single-particle band more dispersive whereas increasing the single-particle band gap suppresses band mixing between the partially filled lower band and the upper band.  $W\rightarrow 0$ corresponds to a perfectly flat band and $\Delta_s\rightarrow\infty$ leads to a single band at low filling, a limit discussed in Ref.~\cite{lin:2014}.  The work presented here considers a more physical model with experimentally realistic numbers for $W$ and $\Delta_s$.  The lobe shows that the Wigner crystal with emergent Luttinger liquid properties found in the approximate model of Ref.~\cite{lin:2014} remains stable and is adiabatically connected to the Wigner crystal in the physical model considered here.
 }
\end{figure}

Recent work \cite{lin:2014} modeling fermions in one dimensional optical lattices with a quantum flat band defined by SOC shows that they can be described with an emergent Luttinger liquid theory \cite{huber:2010,lin:2014} that contrasts with ordinary Luttinger liquid theory \cite{haldane:1981,fisher:1997,miranda:2003,gimarchy:2013}.  In an emergent Luttinger liquid the fermions experience an effective band (generated entirely by interactions) in which Luttinger liquid-like properties appear from the interaction alone.  The emergent Luttinger liquid theory (and numerical diagonalization) showed that the ground state of the system is a Wigner crystal of spinors.  The low energy excitations of the crystal displayed emergent kinetics and fractionalized charge.  The ground and excited states stemmed from just the $s$-wave interaction that was effectively extended in range because the single-particle basis states (Wannier functions) where elongated.

In this work we build on the results of Ref.~\cite{lin:2014} to model a more realistic Hamiltonian to test the robustness of the emergent Luttinger liquid properties.  Ref.~\cite{lin:2014} made a flat band approximation which assumed zero single-particle bandwidth.  It was argued that a small dispersion would not impact the essential properties of the states found in Ref.~\cite{lin:2014}.  Furthermore, a single band was assumed thereby explicitly ruling out the possibility that band mixing would qualitatively change the nature of the states found.  The realistic model we consider here systematically includes both effects (non-zero bandwidth and band mixing from a second band) to explore the robustness of the Wigner crystal with emergent kinetics.

Figure~\ref{fig_schematic} schematically summarizes our findings. Fig.~\ref{fig_schematic} plots the single-particle band gap, $\Delta_s$, versus the single-particle bandwidth, $W$, for a one-dimensional optical lattice in the presence of SOC.   The slope of a straight line in this plane quantifies the band flatness ratio \cite{zhang:2013}  ($F\equiv \Delta_s/W$).  The lobe in Fig.~\ref{fig_schematic} plots the regime where we find, in this work, that the Wigner crystal is stable and can be described by an emergent Luttinger liquid theory.  In the far right part of the graph, the highly dispersive band favors particles nesting in band minima.  Here a conventional Luttinger liquid theory applies.  In the upper left corner of the diagram, the Wigner crystal destabilizes because the single-particle basis states do not overlap and the nearest neighbor interaction stemming from overlapping Wannier functions vanishes.  Here the interactions cannot lift the degeneracy of the lowest flat band.

We find that the spinor Wigner crystal with emergent kinetics survives realistic effects expected in an optical lattice experiment: a non-zero bandwidth and band mixing.  Studying spectra within the lobe reveals that emergent dispersive states are adiabatically connected to states found in the  approximate model studied in Ref.~\cite{lin:2014}.  We use numerical exact diagonalization to map out the phase diagram and rigorously quantify the location of the lobe for various flatness ratios.  We also find that (within the lobe) band mixing lowers the gap of the Wigner crystal making it less stable than predicted in Ref.~\cite{lin:2014}.  We verify that realistic trapping parameters for a common example atom, $^{40}$K, still allow a Wigner crystal near the trap center even without assuming that a Feshbach resonance can increase the interaction strength.

The outline of the paper is as follows: In Sec.~\ref{sec_model} we construct the full physical model and relate it to the flat-band projected model studied previously in Ref.~\cite{lin:2014}.  We analytically solve the single-particle part of the Hamiltonian to construct the basis in which we represent the full interacting Hamiltonian.  In Sec.~\ref{sec_dispersion_mixing} we study the impact of non-zero bandwidth and band mixing by diagonalizing an interacting model that extrapolates between the single-band projected model and the full physical model.  In Sec.~\ref{sec_phase_diagram} we  map out the phase diagram by numerically diagonalizing the full physical model.  We find a sizable region of stability for the Wigner crystal.  We summarize in Sec.~\ref{sec_Summary}.

\section{Model}
\label{sec_model}

In this section we derive a tight-binding model of $N$ fermions in a one-dimensional optical lattice with an equal population of two hyperfine states.  We incorporate the lowest (nearly flat) band and the second band.  We solve the single-particle tight-binding limit analytically to obtain the band gap and the bandwidth.  We then derive the tight-binding form of the $s$-wave interaction term.  We include all intra and inter-band interaction terms.   The full model constructed in this section will then be diagonalized in later sections to compare with results reported previously \cite{lin:2014} on the projected flat band model. 

We start with a first-quantized non-interacting Hamiltonian that adds SOC to the optical lattice potential \cite{zhang:2013,lin:2014}:
\[ H_0^{s} = \frac{p_x^2}{2 m} - s E_R \cos^2 (k_L x) + \left( \frac{\hbar k_R}{m}
   \right) p_x \sigma_z + \Omega \sigma_x, \]
where $p_x$ is the momentum of particles of mass $m$, the second term is the optical lattice potential created by counter-propagating lasers with wave
vector $k_L$, and the lattice depth is $s E_R$, where $E_R = \hbar^2 k_L^2 / 2 m$
is the recoil energy.   The third term
describes spin-orbit coupling created by Raman lasers with wave
vector $k_R$, and $\tmmathbf{\sigma}= (\sigma_x, \sigma_y,
\sigma_z)$ are the Pauli spin matrices.  In the last term, $\Omega$
is the Rabi frequency which acts as the Zeeman field strength. In the following, we
choose a lattice spacing $\pi / k_L$ as the length unit. In these units, 
$k_R = \pi / 2$ implies $k_R = k_L / 2$.

Here we have assumed a quasi-one dimensional limit derived from strong trapping along the perpendicular ($y$ and $z$) directions.  The particles are only allowed to propagate along $x$.  The primary effect is to renormalize the $s$-wave scattering length.  We incorporate the effect of trapping along perpendicular directions when we estimate experimental parameters in the last section. 

  To pass to the tight binding limit we rewrite the Hamiltonian in second quantized form \cite{lin:2014}: 
   \begin{equation}
  H_0 = - 2 t \sum_{k,\sigma} \cos (k + k_R \sigma) c_{k \sigma}^{\dagger}
  c_{k \sigma}^{\vphantom{\dagger}} + \Omega \sum_{k,\sigma \neq  \bar{\sigma}} c_{k \sigma}^{\dagger} c_{k
  \bar{\sigma}}^{\vphantom{\dagger}}, \label{eq_tbH}
\end{equation}
where  $c_{k,\sigma}^{\dagger}$ creates a fermion at wavevector $k$ in one of two hyperfine states with pseudo-spin indices $\sigma = \uparrow,
\downarrow$, and $t$ is the nearest neighbor hopping matrix element.  We set $k_R \sigma = \pm k_R$ for $\sigma = \uparrow,
\downarrow$ respectively.  We have checked, by directly solving the continuum model, that the tight binding model presented here reproduces the band energies of the continuum model to within $5\%$ for the parameters we study.

Equation~\eqref{eq_tbH} can be solved analytically by passing to the band basis, labeled by $\alpha = +, -$.  The appendix shows that a unitary transformation leads to a diagonal form: 
\begin{equation}
  H_0 = \sum_{k,\alpha} E_{\alpha} (k) \chi_{k \alpha}^{\dagger} \chi_{k
  \alpha}^{\vphantom{\dagger}}, \label{eq_H_0}
\end{equation}
with eigenvalues:
\begin{equation} 
E_{\pm} (k) = - 2 t \cos k \cos k_R \pm \sqrt{\Omega^2 + 4 t^2 \sin^2 k
   \sin^2 k_R}. \label{eq_band}
    \end{equation}
The eigenvectors are:
$
\chi_{k \alpha}^{\vphantom{\dagger}} = \sum_{\sigma} v_{k \alpha, \sigma}^{\ast} c_{k \sigma}^{\vphantom{\dagger}},  
$
with coefficients 
 $v_{k +} = \left[ \begin{array}{ll}
  \cos{(\theta_k/2)} & \sin{(\theta_k/2)}
\end{array} \right]^T$ and $v_{k -} = \left[ \begin{array}{ll}
  \sin{(\theta_k/2)} & -\cos{(\theta_k/2)}
\end{array} \right]^T$ where $\cos( \theta_k) = 2 t \sin (k) \sin (k_R) / \sqrt{\Omega^2 + 4 t^2 \sin^2 k
   \sin^2 k_R}$.  The basis states $\chi_{k\alpha}$ define spinors with a magnetic moment orientation that depends on $\theta_k$.

\begin{figure}
  \includegraphics[width=1.0\columnwidth]{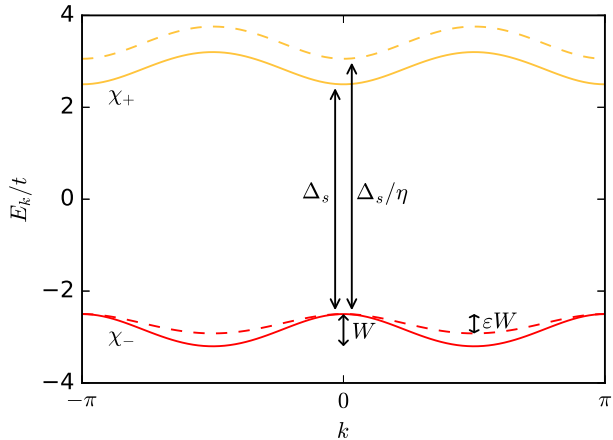}
  \caption{  Single-particle energies as a function of wavevector for the two bands, $\alpha=\pm$.  The solid lines plot Eq.~\eqref{eq_band} for $\Omega = 2.5 t$ and $k_R = k_L / 2$ which lead to a flatness ratio $F \approx 7$.  The dashed lines plot the same but for Eqs.~\eqref{eq_Eeps} and ~\eqref{eq_Eeta} with the dimensionless parameters $\varepsilon=0.6$ and $\eta=0.9$ introduced to tune the single-particle bandwidth and the band gap, respectively.
  }
  \label{fig_sband}
\end{figure}

Figure~\ref{fig_sband} plots the band structure defined by Eq.~\eqref{eq_band}  
in the case of maximal spin-orbital
coupling $k_R = k_L / 2$.  Here we see that the lowest of the two bands is very flat, $F\approx7$.   We quantify the band flatness ratio here using the single-particle bandwidth: 

\begin{eqnarray*}
W = \sqrt{\Omega^2 + 4 t^2} - | \Omega |,
\label{eq_bandwidth}
\end{eqnarray*}
 and the single-particle band gap: 
 \begin{eqnarray*}
 \Delta_s = 2 | \Omega |.
 \label{eq_bandgap}
 \end{eqnarray*}
 
We now use the single-particle basis to represent the inter-atom interaction term.  We consider $s$-wave scattering between atoms.  The interaction in the basis before application of spin-orbit coupling leads to the usual Hubbard interaction between atoms:
\[ H_{\tmop{int}} = \frac{U}{2} \sum_{i, \sigma\neq  \bar{\sigma}} c_{i \sigma}^{\dagger} c_{i
   \bar{\sigma}}^{\dagger} c_{i \bar{\sigma}}^{\vphantom{\dagger}} c_{i \sigma}^{\vphantom{\dagger}}. \]
This interaction is purely onsite because of the local nature of the Wannier functions (before the application of spin-orbit coupling).  After a Fourier transform to momentum space the Hubbard interaction becomes:
\[ H_{\tmop{int}} = \sum_{\{ k \},\sigma\neq  \bar{\sigma}} V_{\{ k \}} c_{k_4 \sigma}^{\dagger}
   c_{k_3 \bar{\sigma}}^{\dagger} c_{k_2 \bar{\sigma}}^{\vphantom{\dagger}} c_{k_1 \sigma}^{\vphantom{\dagger}}, \]
where $V_{\{ k \}}  = (U/2 L) \delta^{'}_{k_4
+ k_3 = k_2 + k_1}$, $\delta'$ indicates momentum
conservation up to multiples of the reciprocal lattice vector, and $L$ is the number
of sites.

The application of spin-orbit coupling has a drastic effect on the single-particle basis states.  The basis states can, for low to intermediate $F$, elongate in real space and overlap between nearest neighbors.  We incorporate spin-orbit coupling by rewriting the interaction in terms of single-particle eigenstates of Eq.~\eqref{eq_tbH}:
\begin{equation}
  H_{\tmop{int}} = \sum_{\{ k, \alpha \}} \tilde{V}_{\{ k \alpha \}} \chi_{k_4
  \alpha_4}^{\dagger} \chi_{k_3 \alpha_3}^{\dagger} \chi_{k_2 \alpha_2}^{\vphantom{\dagger}}
  \chi_{k_1 \alpha_1}^{\vphantom{\dagger}}, \label{eq_H_int}
\end{equation}
where the interaction matrix elements, $\tilde{V}_{\{ k, \alpha \}} = (U/2 L)  \sum_{\sigma\neq  \bar{\sigma}} v_{k_4
\alpha_4, \sigma}^{\ast} v_{k_3 \alpha_3, \bar{\sigma}}^{\ast} v_{k_2
\alpha_2, \bar{\sigma}} v_{k_1 \alpha_1, \sigma}  \allowbreak \delta'_{k_4 + k_3 =
k_2 + k_1}$, incorporate both the interaction and spin-orbit coupling.  

\begin{figure}
  \includegraphics[width=1.0\columnwidth]{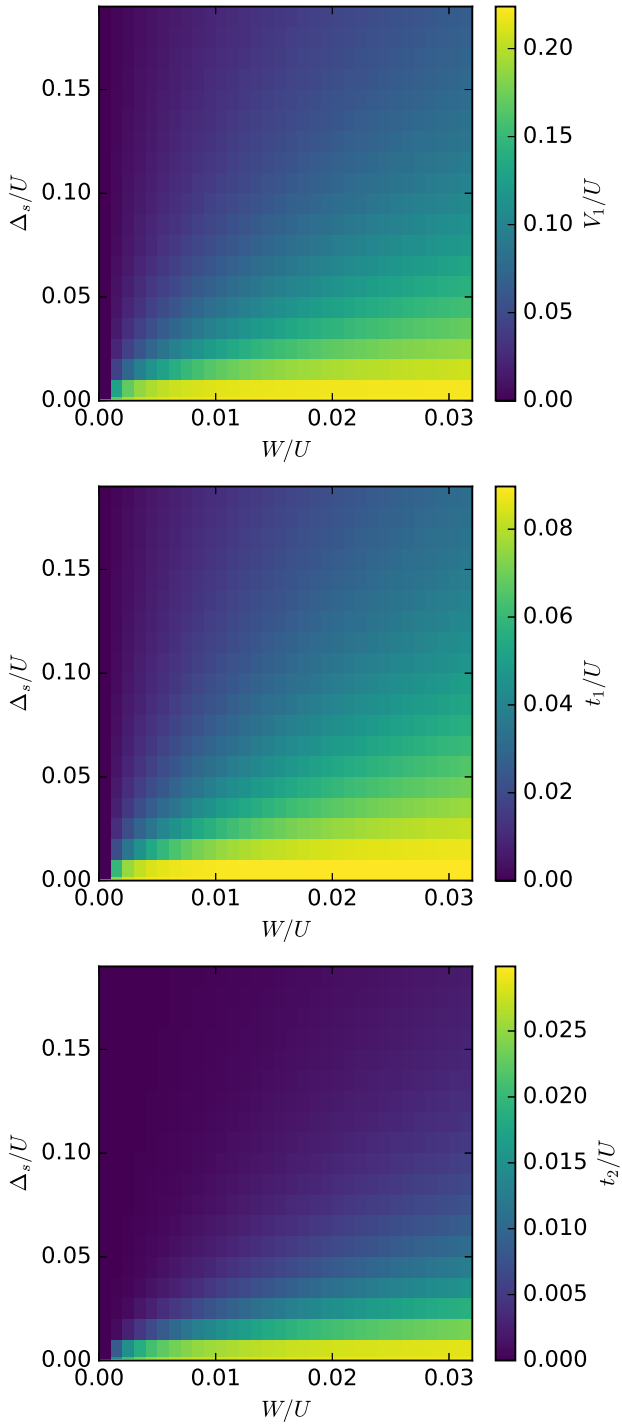}
  \caption{ Strength of leading interaction matrix elements in the Wannier basis plotted as a function of both the single-particle band gap $\Delta_s$ and the bandwidth $W$. The top, middle, and bottom panels correspond to coefficients of the nearest neighbor density-density interaction $V_1\chi_{i,-}^{\dagger}\chi_{i,-}\chi_{i+1,-}^{\dagger}\chi_{i+1,-}$, next-nearest neighbor density assisted tunneling $ -t_1\chi_{i+2,-}^{\dagger}\chi_{i+1,-}^{\dagger} \chi_{i+1,-}\chi_{i,-}$, and the next-next-nearest neighbor density assisted tunneling $ t_2\chi_{i+3,-}^{\dagger}\chi_{i,-}^{\dagger}\chi_{i,-}\chi_{i+1,-}$, respectively.  The size of the coefficients in the centers of the panels (intermediate $\Delta_s$ and $W$) shows that here we expect nearest neighbor correlations to be relevant in a nearly flat band.
  \label{fig_conditional_hopping}}
\end{figure}

Passing back to Wannier functions in real space one can see that, for low to intermediate $F$, the Wannier functions have been considerably elongated by spin-orbit coupling \cite{lin:2014} to overlap in neighboring sites.  To see the impact of elongation on the tight-binding parameters for interaction terms Fig.~\ref{fig_conditional_hopping} plots the lowest-band coefficients for the nearest-neighbor density-density interaction ($V_1$), the next-nearest neighbor density assisted tunneling ($t_1$), and the next-next-nearest neighbor density assisted tunneling ($t_2$) as a function of the band width and the single-particle band gap.  Here we see that, for intermediate $\Delta_s$ and $W$, the nearest neighbor interaction can become sizable. We must therefore include nearest-neighbor interaction terms when writing the interaction in the $\chi$ basis.  
 
Including interactions, the total Hamiltonian becomes:
\begin{equation}
  H = H_0 + H_{\tmop{int}} \label{eq_fullH},
\end{equation}
where $H_0$ is diagonal in the $\chi$ basis, Eq.~\eqref{eq_H_0}.  The interaction term is off-diagonal and, for certain parameters, yields a formidable non-perturbative problem because the lowest band becomes nearly degenerate. The study of $H$ will form the focus of the rest of the paper.  

Reference~\cite{lin:2014} used a flat band approximation to study Eq.~\eqref{eq_fullH} for $N/L=1/2$.  In the flat band approximation, two limits were taken.  First, all particles are projected onto the lowest band, $\alpha=-$.  At partial filling, lowest band projection can be thought of as setting $\Delta_s\rightarrow\infty$.  Second, the single-particle dispersion was assumed to be irrelevant and $H_0$ was dropped.  In this approximation, the projected Hamiltonian becomes:
\begin{equation}
  H^P = \sum_{\{ k \}} V^P_{\{ k \}} \chi_{k_4 -}^{\dagger} \chi_{k_3
  -}^{\dagger} \chi_{k_2 -}^{\vphantom{\dagger}} \chi_{k_1 -}^{\vphantom{\dagger}}, \label{eq_projH}
\end{equation}
where $V^P_{\{ k \}} = (U/L) v_{k_4 -, \uparrow}^{\ast} v_{k_3 -,
\downarrow}^{\ast} v_{k_2 -, \downarrow} v_{k_1 -, \uparrow} \delta'_{k_4 +
k_3 = k_2 + k_1}$.  

We see explicitly that $H^P$ defines a non-perturbative problem because there are no other terms in the model.  The flat band approximation assumes that inclusion of single-particle terms ($H_0$) will merely perturb the physics found by diagonalizing Eq.~\eqref{eq_projH} while the low energy eigenstates remain in the same universality class.  We test the flat band approximation by comparing solutions to Eqs.~\eqref{eq_fullH} and \eqref{eq_projH}.   Non-zero bandwidth and mixing due to interaction effects should perturb the low energy eigenstates.  We consider the impact of both finite bandwidth and inter-band mixing in the following.  

\section{Band Dispersion and Band Mixing}
\label{sec_dispersion_mixing}

\begin{figure}
  \includegraphics[width=1.0\columnwidth]{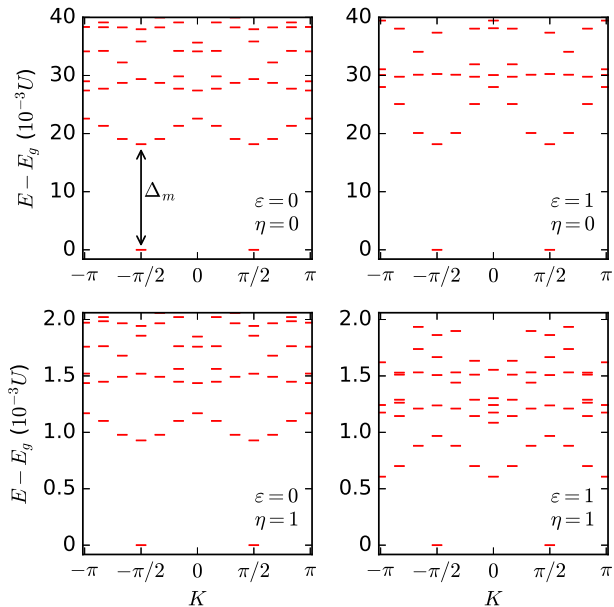}
  \caption{  Many-body energies versus total wavevector obtained from diagonalizing Eq.~\eqref{eq_Hlimits} in four distinct limits of the parameters $\varepsilon$ and $\eta$.  The energy zero is the ground state energy, $E_g$. The projected model, Eq.~\eqref{eq_projH}, is retrieved for $\varepsilon=\eta=0$ (upper left).  A non-zero bandwidth is introduced for $\varepsilon=1$ and $\eta=0$ (upper right) while a second flat band is introduced for $\eta=1$ and $\varepsilon=0$ (lower left).  The full physical model, Eq.~\eqref{eq_fullH}, is retrieved for $\varepsilon=\eta=1$ (lower right).  
 The ground state remains a Wigner crystal with 2-fold sublattice 
  degeneracy in all four panels and we have checked that the ground state
  energy and ground state wave functions are adiabatically connected between the four limits.  Comparing the $\eta=0$ to $\eta=1$ cases shows that band mixing lowers the many-body band gap, $\Delta_m$, by a factor of $\approx 20$.  We have used the following parameters: 
  $N = 6$, $L = 12$, $t = 0.01 U$, $\Omega = 0.025 U$, $k_R = k_L / 2$, i.e., $\Delta_{s} = 0.05 U$,
  and $W = 0.007 U$.  This corresponds to $F \approx 7$.   \label{fig_four_spectrum}}
\end{figure}

In this section we use exact diagonalization to study the impact of finite bandwidth and band mixing separately.  We introduce tuning parameters to $H$ so we can extrapolate between $H$ and $H^P$ to thus allow separate analyses of each effect.  By examining the spectrum and computing eigenstate overlaps we find that band mixing alone lowers the gap between the ground and first excited state by a factor of at least $\approx$ 20.  When we include both band mixing and non-zero bandwidth we also find that the many-body dispersion shifts in wave-vector.

We start by inserting tuning parameters into the single-particle energy to allow a separation of effects.  For the lowest band we tune the width of the lowest band using an artificial tuning parameter, $\varepsilon$:
\begin{equation} 
E_- (\varepsilon, k) \equiv - \Omega + [E_- (k) + \Omega ]\varepsilon,
\label{eq_Eeps}
\end{equation}
while for the second band we tune the band gap with $\eta$:
\begin{equation} 
E_+ (\eta, k) \equiv E_+ (k) / \eta,
\label{eq_Eeta}
\end{equation}
The dashed lines in Fig.~\ref{fig_sband} show the effect of the parameters $\varepsilon$ and $\eta$.  For $\varepsilon=0$ we recover the flat band limit and for $\eta=0$ we set the band gap to infinity to recover the single band limit.  The limit  $\varepsilon=\eta=1$ returns us to the physical single-particle energy, Eq.~\eqref{eq_band}.

\begin{figure}
  \includegraphics[width=1.0\columnwidth]{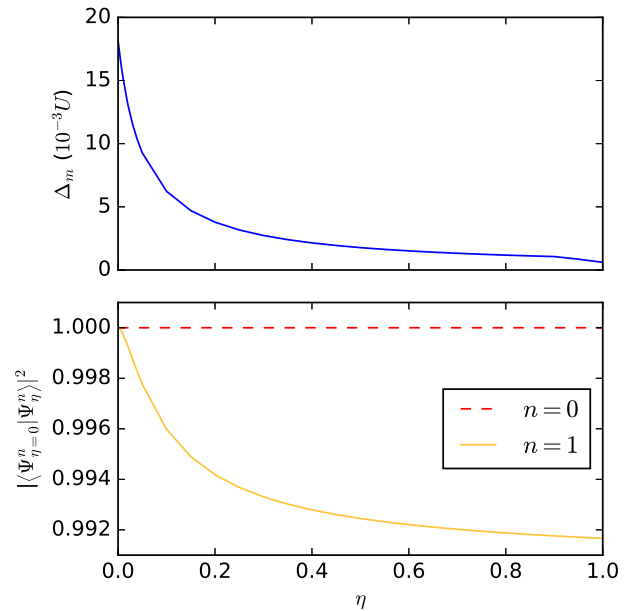}
  \caption{The many-body gap (top) and many-body wavefunction overlap (bottom) obtained from diagonalizing Eq.~\eqref{eq_Hlimits} as a function of the dimensionless parameter  $\eta$.  Here we see that the many-body gap is significantly lowered as we introduce a second single-particle band by increasing $\eta$.  The bottom panel plots the overlap between the $\eta=0$ wavefunction and the wavefunction for $\eta\geq0$ for both the ground state ($n=0$) and the first excited state ($n=1$) to show that the second single-particle band alters the nature of just the first excited state.  The parameters are the same as Fig.~\ref{fig_four_spectrum} but for $\varepsilon=1$
  \label{fig_overlap}
 }
\end{figure}

By adding interactions we construct a model that allows us to tune between different limits:
\begin{equation}
H_{\varepsilon,\eta}=\sum_{k} \left [
E_{-} (\varepsilon,k) \chi_{k -}^{\dagger} \chi_{k
  -}^{\vphantom{\dagger}} 
  +E_{+} (\eta,k) \chi_{k +}^{\dagger} \chi_{k
  +}^{\vphantom{\dagger}}  \right ]
  +H_{\tmop{int}}, 
  \label{eq_Hlimits}
\end{equation}
For $\varepsilon\rightarrow0$ and  $\eta\rightarrow0$ we have, at partial filling, the flat single-band limit: $\underset{\varepsilon,\eta\rightarrow0}{\lim} H_{\varepsilon,\eta}=H^P$ and for 
$\varepsilon=\eta=1$ we have the full physical model $ H_{\varepsilon=1,\eta=1}=H$.  We stress that $\varepsilon$ and $\eta$ are unphysical tuning parameters that are designed to test eigenstate adiabaticity between two physical limits: $\varepsilon=\eta=1$ and  $\varepsilon=\eta=0$.

We diagonalize Eq.~\eqref{eq_Hlimits} in different limits to explore the impact of single-particle band effects on interaction-driven physics.  Fig.~\ref{fig_four_spectrum} shows the results of diagonalizing Eq.~\eqref{eq_Hlimits} in four different limits.  The top left panel reproduces the results found in Ref.~\cite{lin:2014} for the flat single-band model, $H^P$.  Here we see that the lowest energy state is two-fold degenerate and corresponds to a Wigner crystal of spinors that can be generated by just the diagonal density-density interaction term in Eq.~\eqref{eq_projH}.  The two degenerate states arise because of  the sublattice degeneracy for the two ways of placing the crystal on the one-dimensional lattice.  There is a gap to the lowest band of excitations.  Ref.~\cite{lin:2014} pointed out that these states show emergent kinetics due to the finite many-body bandwidth driven entirely by off-diagonal terms in  Eq.~\eqref{eq_projH}.  The focus of our work here is to probe the stability of this low-energy structure  
as we introduce a second band and allow non-zero bandwidth. 

\begin{figure}
  \includegraphics[width=1.0\columnwidth]{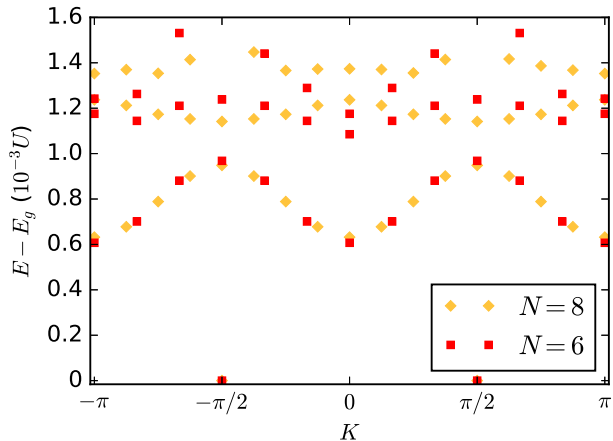}
  \caption{ The same as Fig.~\ref{fig_four_spectrum} but for the full physical model, Eq.~\eqref{eq_fullH}, where the squares (diamonds) are for $N=6$ ($N=8$) particles on $L=12$ ($L=16$) sites.   The data collapse shows that the ground and first excited states are already in the thermodynamic limit.
  \label{fig_data_collapse}}
\end{figure}

The top-right and bottom-left panels of  Fig.~\ref{fig_four_spectrum} show the result of adding finite bandwidth ($\varepsilon=1$) and band mixing ($\eta=1$), respectively.  Here we see that setting $\varepsilon=1$ does very little to the many-body spectrum at low energies.  For $F=7$ the band is so flat that the small but finite single-particle dispersion does not perturb the large interaction much.  But for $\varepsilon=0$ and $\eta=1$  we see that bringing two flat bands relatively near each other causes the many-body gap, $\Delta_m$, to decrease by a factor of $\approx20$ while keeping the structure of the low energy states qualitatively the same.

The bottom-right panel of Fig.~\ref{fig_four_spectrum} shows the spectrum for the full model, $H$.  Here we see that including both finite bandwidth and band mixing not only lowers the gap appreciably but the many-body excited states are shifted in $K$-space so that the many-body dispersion has a minimum at $K=0$ instead of $K=\pm \pi/2$. Here the non-zero single-particle dispersion mixed the lowest energy many-body excited states. Otherwise the qualitative features of the low energy states remains the same as we go from $\varepsilon=\eta=0$ to $\varepsilon=\eta=1$.

The top panel of Fig.~\ref{fig_overlap} shows the decrease in the many-body gap as the single-particle band gap is lowered.  Here we keep a non-zero single-particle dispersion ($\varepsilon=1$) but we tune the single-particle gap from infinity to $\Delta_s$.  The gap never drops to zero thus signaling that the low energy states in the full Hamiltonian, $H$, are adiabatically connected to the those of the projected Hamiltonian, $H^P$.  

The mixing of the many-body excited states drives the lowering of the gap.  To see this we plot the overlap of the lowest two many-body states in the lowest panel of Fig.~\ref{fig_overlap}.  Here we see that the ground state remains unperturbed but the mixing of the excited states somewhat lowers the overlaps from the single-band ($\eta=0$) limit.  Nonetheless we see that the overlap remains large and does not show any cusps.  There are therefore no transitions as we lower the band gap for $F=7$.  In the following sections we will vary $F$ to find transitions (where the many-body gap vanishes).

\begin{figure}
  \includegraphics[width=1.0\columnwidth]{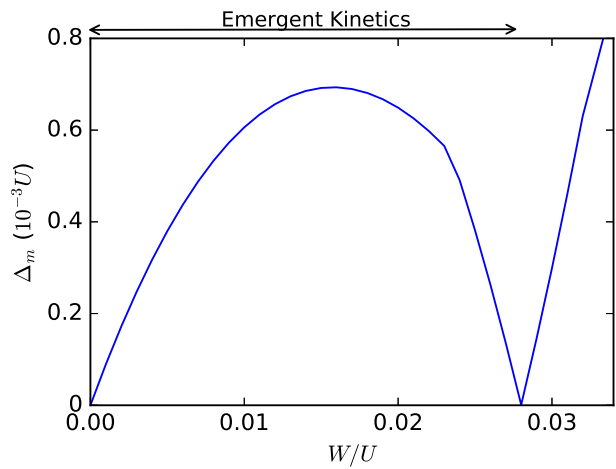}
  \caption{ The many-body energy gap plotted as function of bandwidth.  The parameters are the same as Fig.~\ref{fig_four_spectrum} but for the full physical model, Eq.~\eqref{eq_fullH}, with the band gap held constant, $\Delta_s = 0.08 U$.  Here we see that at zero bandwidth the single-particle basis states have no spread and, as a result, the interaction remains onsite and cannot lift the degeneracy.  But as the bandwidth increases, the nearest-neighbor interaction terms lift the degeneracy to reveal the Wigner crystal ground state and opens a gap to a set of emergent excitations captured by an effective Luttinger liquid theory.  But as the bandwidth increases further the gap closes as the Wigner crystal transitions to a conventional Luttinger liquid regime.  
   \label{fig_gap_reopening}}
\end{figure}

We have checked that our results presented here do not change as we increase particle number and are therefore valid in the thermodynamic limit.   Fig.~\ref{fig_data_collapse} shows data collapse in the spectrum.  The low energy states fall on one another indicating a consistency in scaling to the thermodynamic limit.  This was also found for $H^P$ in Ref.~\cite{lin:2014} further showing that the low energy eigenstates of both $H$ and $H^P$ are in the same universality class.

\section{Phase Diagram and Stability}
\label{sec_phase_diagram}

We now map out the stability phase diagram of the interaction-only spinor Wigner crystal phase of Eq.~\eqref{eq_fullH}.  Instabilities arise as we increase the single-particle bandwidth.  For large $W$ the particles gain in energy by nesting in the single-particle band minima.  There is therefore a transition from the interaction-dominated regime (with emergent kinetics) to a weakly interacting state (a conventional Luttinger liquid) as the bandwidth is increased.  Increasing the single-particle band gap also drives a transition.  At first we expect that increasing $\Delta_s$ might favor the approximation that led to emergent kinetics.  But note that large $F$ implies that the lowest-band Wannier functions have little overlap between nearest neighbor sites \cite{lin:2014}.  As a result, increasing $F$ decreases density assisted hopping terms between neighbors (see Fig.~\ref{fig_conditional_hopping}) and therefore suppresses emergent kinetics.  We thus expect a transition to a gapless regime as $\Delta_s$ and therefore $F$ is increased. 

\begin{figure}
  \includegraphics[width=0.8\columnwidth]{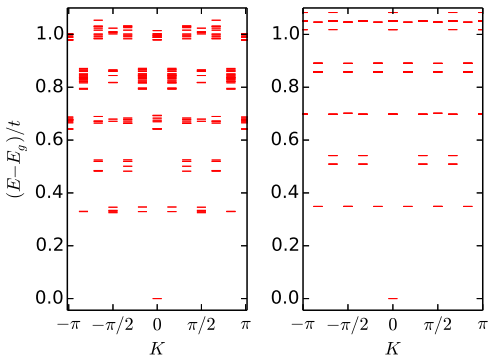}
  \caption{Characteristic many-body spectrum of Eq.~\eqref{eq_fullH} computed for a weakly interacting case (left panel, $U=t/2$) and the non-interacting case (right panel, $U=0$).  We have also set $N=6$, $L=12$, $\Omega=2.5t$, and $k_R = k_L / 2$.  These parameters lead to a flatness ratio used in the other figures as well, $F\approx7$. A comparison of both panels shows that the spectra are qualitatively similar, i.e., states occur at the same wavevectors and nearly the same energies.  We can therefore think of the ground state in both cases as a partially filled band of weakly interacting fermions.  The weakly interacting case conforms to conventional Luttinger liquid theory.}
  \label{fig_filling_band}
\end{figure}

We increase $W$ and diagonalize Eq.~\eqref{eq_fullH} to find the lowest energy eigenstates. Note that increasing $W$ impacts $H_0$ directly and $H_{\tmop{int}}$ indirectly through the change in basis states $ \chi_{k \alpha}$.   Fig.~\ref{fig_gap_reopening} plots the many-body gap as a function of the bandwidth.  We see that the many-body gap starts from zero at $W=0$.  For $W\rightarrow0$ we have $F\rightarrow\infty$ and Eq.~\eqref{eq_projH} is a good approximation to Eq.~\eqref{eq_fullH}.  But in this limit the are essentially no nearest neighbor terms to lift the massive degeneracy of spinless particles in the lowest flat band.  Here the flat band remains gapless.  As we increase the $W$ term, nearest neighbor interaction terms (not single-particle terms) drive the formation of a spinor Wigner crystal with emergent kinetics and the many-body gap opens.

\begin{figure}
  \includegraphics[width=1.0\columnwidth]{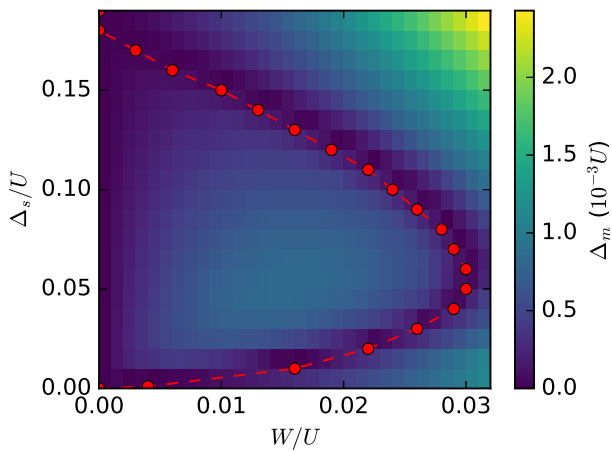}
  \caption{ Stability phase diagram of the Wigner crystal with emergent kinetics plotted as function of both the single-particle band gap and the bandwidth.  The color coding plots the size of the many-body gap obtained from diagonalization of Eq.~\eqref{eq_fullH} for $N=6$, $L=12$, and $k_R = k_L / 2$.  The circles plot the points where the many-body gap vanishes and the line is a guide to the eye.  The Wigner crystal is stable within the lobe.  Outside the lobe we have a conventional Luttinger liquid with a gap set by finite-size effects. 
  \label{fig_pd}}
\end{figure}

Upon increasing $W$ further the many-body gap closes and a new state arises in Fig.~\ref{fig_gap_reopening}.  Here the Wigner crystal destabilizes to a more conventional state where $H_0$ and  interactions compete in Eq.~\eqref{eq_fullH}.   Conventional Luttinger liquid theory can be used to show that the particles tend to sit about the single-particle band minimum.  The ground state in the large $W$ regime can be understood by filling the lowest single-particle band with weakly interacting fermions.  Characteristic spectra that arise for large $W$ are shown in the left panel of Fig.~\ref{fig_filling_band}.  The right panel shows that non-interacting spectra give nearly the same results.  In both panels the gaps are due to finite-size effects and there is no ground state degeneracy since filling of the lowest single-particle band leads to a unique $K$.  We can therefore understand the large $W$ limit in a weakly interacting picture of band filling of spinless fermions.

We culminate our findings in a phase diagram that plots the stability of the Wigner crystal and its emergent kinetics.  The shading in Fig.~\ref{fig_pd} plots the size of the many-body gap as a function of both the single-particle bandwidth and band gap.  The circles denote critical points where the many-body gap closes and the ground state degeneracy changes from two (Wigner crystal with emergent kinetics) to one (conventional Luttinger liquid regime).  Inside the lobe nearest neighbor interactions establish the many-body gap but outside the lobe the gap is, for our finite size simulations, set by the finite size of the system.

The parameters needed to reach the central part of the lobe are accessible with current experiments.   We assume $^{40}$K atoms with two hyperfine levels populated to define the pseudospin.  To compute the tight-binding parameters we solve the periodic Schr\"odinger equation using Mathieu functions and compute the Wannier functions in the usual way \cite{jaksch:1998,hui:2016}.  We find that for a perpendicular confinement of $60E_R$, a lattice depth of $s\approx 13$, and a bare scattering length of $a_s=104a_0$ (where
$a_0$ is the Bohr radius) we can achieve $t=0.01U$, where $t\approx
0.01E_R$ and $U\approx E_R$ .  The Zeeman field can then be chosen to be
$\Omega\approx 0.025E_R$ with $k_R=k_L/2$.  This leads to $F\approx 7$ and corresponds to a central part of the lobe in Fig.~\ref{fig_pd} with $\Delta_s/U=0.05$ and $W/U=0.007$.

The parabolic trapping potential competes with the many-body gap to limit the size of the Wigner crystal near the trap center.  We can estimate the size of the Wigner crystal by equating the energy cost required to overcome $\Delta_m$ with the trapping potential energy.  The trapping potential is $m \omega_{\tmop{tr}}^2 x /2$, where $\omega_{\tmop{tr}}$ is the trapping frequency.   We estimate the position $x_{max}$ where the
crystal no longer exists using $\Delta_m = m\omega_{\tmop{tr}}^2 x_{max} /2$.  For $^{40}$K on a lattice formed by
lasers with wavelength $826$ nm and a realistic trapping strength
$\omega_{\tmop{tr}}=40-70$Hz we find a crystal size of $2x_{max}\approx 20-34$ lattice sites.  This size was estimated using $\Delta_m$ from the full Hamiltonian, Eq.~\eqref{eq_fullH}.  $x_{max}$ is smaller than the size estimated using the single-band projected model (We find $\approx100$ sites with Eq.~\eqref{eq_projH} \cite{lin:2014}) because the band mixing lowered $\Delta_m$.  Nonetheless we find that band mixing and finite bandwidth in $H$ still allow a Wigner crystal with emergent kinetics in a small region near the center of the trap.  The strength of the crystal can be increased by increasing the strength of the interaction (and therefore $U$) using a Feshbach resonance.  

\section{Summary}
\label{sec_Summary}

We have studied a model of two-component fermionic atoms in a one-dimensional optical lattice in the presence of SOC.  We have mapped out the stability phase diagram of a spinor Wigner crystal with emergent kinetics in its low energy excitation state space.  Our results demonstrate the parameter regime of validity of the approximations made in Ref.~\cite{lin:2014} by showing that the projected approximate model, Eq.~\eqref{eq_projH}, captures the essential properties of the low energy states of the full model, Eq.~\eqref{eq_fullH}.  We find that band mixing lowers the gap of the Wigner crystal by at least a factor of $\approx20$.  Band mixing and a finite bandwidth also shift the low energy momenta of the emergent many-body models from a total momentum of $K=\pm\pi/2$ in approximate case, Eq.~\eqref{eq_projH}, to $K=0$ in the full model, Eq.~\eqref{eq_fullH}.  Nonetheless, the Wigner crystal and its emergent modes show sufficient stability to occupy the central region of a trapped optical lattice experiment.  We estimate $\approx 30$ sites for the bare interaction between $^{40}$K atoms in a trap.  A Feshbach resonance can be used to increase the strength of the states discussed here.

\begin{acknowledgments}
We acknowledge helpful discussions with C. Zhang and support from AFOSR (FA9550-15-1-0445)
and ARO (W911NF-16-1-0182). 
\end{acknowledgments}

\bibliographystyle{apsrev4-1}\bibliography{references.bib}

\appendix

\section{Mapping Between Spin and Band Operators}\label{sec_appendix}

In this section we detail the mapping between fermions in the spin basis [Eq.~\eqref{eq_tbH} in terms of $c_{k \sigma}$] and fermions in the band basis [Eq.~\eqref{eq_H_0} in terms of $\chi_{k \alpha}$].  We start by rewriting the single-particle tight-binding model, Eq.~\eqref{eq_tbH}, in matrix form:
\begin{eqnarray*}
  H_0 & = & - 2 t \sum_{k,\sigma} \cos (k + k_R \sigma) c_{k \sigma}^{\dagger}
  c_{k \sigma}^{\vphantom{\dagger}} + \Omega \sum_{k,\sigma\neq \bar{\sigma}} c_{k \sigma}^{\dagger} c_{k
  \bar{\sigma}}^{\vphantom{\dagger}}\\
   & = & C_k^{\dagger} [h_0 (k) I +\mathbf{h} (k) \cdot \tmmathbf{\sigma}] C_k^{\vphantom{\dagger}},
\end{eqnarray*}
where $I$ is the identity matrix, $C_k = \left( \begin{array}{ll}
  c_{k \uparrow} & c_{k \downarrow}
\end{array} \right)^T$, $h_0 (k) = - 2 t \cos k \cos k_R$, and $\mathbf{h} (k) =
(\Omega, 0, 2 t \sin k \sin k_R)$.  We can rewrite $\mathbf{h} (k)$ in spherical coordinates:

\[ \mathbf{h} (k) = h_k (\sin \theta_k \cos \phi_k, \sin
   \theta_k \sin \phi_k, \cos \theta_k), \]
where $h_k$ is the magnitude, $\theta_k$ is the polar angle, $\phi_k$ is the azimuthal angle. In the case studied here we have 
$\cos \theta_k = 2 t \sin k \sin k_R / h_k$ and $\phi_k = 0$.

We can now diagonalize eigenvalues $H_0$ to obtain the eigenvalues $E_{\pm}(k)$ and eigenvectors $v_{k
\pm}$ using a unitary transformation:
\[ U^{\dagger} (k) [h_0 (k) I +\mathbf{h} (k) \cdot \tmmathbf{\sigma}] U (k) =
   \tmop{diag} \{ E_{+}(k), E_{-}(k) \}, \]
where we find:
\[ E_{\pm}(k) = h_0 (k) \pm | \mathbf{h} (k) |, \]
with:
\[ U (k) = \left( \begin{array}{ll}
     v_{k +} & v_{k -}
   \end{array} \right) = \left( \begin{array}{ll}
     \cos (\theta_k / 2) & \sin (\theta_k / 2)\\
     \sin (\theta_k / 2) & - \cos (\theta_k / 2)
   \end{array} \right). \]

We can then use the unitary transform to define the band operators $\chi_{k \pm}$:

\[ C_k = U (k) \Chi_k \]
where $\Chi_k = \left( \begin{array}{ll}
  \chi_{k +} & \chi_{k -}
\end{array} \right)^T$, so that:
\[ H_0 = \sum_k \left[ E_{-}(k) \chi_{k -}^{\dagger} \chi_{k -}^{\vphantom{\dagger}} + E_{+}(k) \chi_{k
   +}^{\dagger} \chi_{k +}^{\vphantom{\dagger}} \right]. \]  This shows that the single-particle tight-binding model, Eq.~\eqref{eq_tbH}, in the spin basis can be diagonalized by rewriting the model in the band basis, Eq.~\eqref{eq_H_0}.

\end{document}